# Ultrahigh-sensitivity optical power monitor for Si photonic circuits

Takaya Ochiai[1], Kei Sumita[1], Shuhei Ohno[1], Stéphane Monfray[2], Frederic Boeuf[2],

Kasidit Toprasertpong[1], Shinichi Takagi[1], Mitsuru Takenaka[1*]

[1] Department of Electrical Engineering and Information Systems, The

University of Tokyo, 7-3-1 Hongo, Bunkyo-ku, Tokyo 113-8656, Japan,

Phone: +81-3-5841-6733, Fax: +81-3-5841-8564,

[2] STMicroelectronics, 850 Rue Jean Monnet 38920 Crolles, France

*E-mail: takenaka@mosfet.t.u-tokyo.ac.jp

**A phototransistor is a promising candidate as an optical power monitor in Si photonic circuits since the internal gain of photocurrent enables high sensitivity. However, state-of-the-art waveguide-coupled phototransistors suffer from a responsivity of lower than $10^3$ A/W, which is insufficient for detecting very low power light. Here, we present a waveguide-coupled phototransistor consisting of an InGaAs ultrathin channel on a Si waveguide working as a gate electrode to increase the responsivity. The Si waveguide gate underneath the InGaAs ultrathin channel enables the effective control of transistor current without optical absorption by the gate metal. As a result, our phototransistor achieved the highest responsivity of approximately $10^6$ A/W among the waveguide-coupled phototransistors, allowing us to detect light of 621 fW propagating in the Si waveguide. The high responsivity and**

**the reasonable response time of approximately 100 μs make our phototransistor promising as an effective optical power monitor in Si photonics circuits.**

The rapid progress in Si photonics over the past decade has made it possible to fabricate large-scale photonic circuits on a Si wafer, which are being widely used as optical interconnection devices in datacenters[1]. In addition to optical fiber communication, many applications such as deep learning[2–4], quantum computing[5–8], and sensing[9,10] are emerging for Si programmable photonic circuits[11], in which numerous optical phase shifters that can control the optical phase of light propagating in a Si waveguide with an electrical signal are integrated. To reconfigure photonic circuits for a specific purpose, it is crucial to accurately set phase values of all phase shifters. Since an initial phase error of a phase shifter is unavoidable owing to variations in device fabrication, accurate initialization and setting a target phase value are critical issues for programmable photonic circuits. Various methods have been proposed to configure phase shifters by monitoring outputs of a photonic circuit[12,13]; however, the simplest and most reliable method is to integrate numerous optical power monitors in a photonic circuit[14–16]. For a non-invasive optical power monitoring in a waveguide, monitoring the change in the conductivity of the waveguide caused by optical absorption through capacitive coupling is one of the promising methods[17]. However, this method requires a relatively high optical input power and phase-sensitive detection with additional electronics for achieving high sensitivity. Ge photodetectors (PDs) widely used in Si photonic circuits are another candidate for use as an optical power monitor[18]; however, the dark current of a Ge PD is rather high, thereby not suitable for detecting very weak light[19,20]. Moreover, an additional photonic circuit for tapping optical power and a transimpedance amplifier are required for detection, and as the number of Ge PDs increases, the complexity of photonic and electrical circuits also increases[21]. Therefore, a high-sensitivity optical power monitor that can be integrated with Si programmable photonic circuits in a simple manner is essential for reconfiguring their functionality accurately.

In this paper, we present an ultrahigh-sensitivity waveguide-coupled phototransistor with an InGaAs ultrathin membrane. Phototransistors based on bipolar junction transistors (BJTs)[22–26], junction field-effect transistors (JFETs) [27,28], and metal-oxide-semiconductor field-effect transistors (MOSFETs)[29–42] can have high photosensitivity owing to the amplification of photocurrent. Although BJT-based phototransistors have been developed so far for high-speed optical communication, their responsivity is typically below $10^3$ A/W because the gain for the photocurrent cannot exceed the gain for the corrector current to the base current. In contrast, MOSFET-based phototransistors called photoFETs can achieve high responsivity. A responsivity exceeding $10^6$ A/W has been reported for surface-illuminated photoFETs. However, a waveguide-coupled photoFET requires a gate electrode near the optical waveguide, which increases optical loss due to optical absorption by the gate metal[29–31]. To prevent optical absorption by the gate metal, the gate electrode should be placed away from the waveguide, losing effective gating[42,43]. As a result, high responsivity has not yet been demonstrated for waveguide-coupled photoFETs.

To resolve this issue, we propose to use a Si waveguide as a back gate for a waveguide-coupled photoFET. As shown in Fig. 1a, an InGaAs ultrathin membrane is bonded onto a Si waveguide with $Al_2O_3$ gate dielectric. A gate voltage can be applied to the InGaAs channel through the p-type Si waveguide, allowing us to eliminate the gate metal that is required for typical waveguide-coupled photoFETs. The effective control of the transistor current flowing in the InGaAs ultrathin channel is achieved while avoiding optical loss due to optical absorption by a gate metal. As a result, we can realize a highly sensitive waveguide-coupled phototransistor. In addition to the superior electrostatic control, with the InGaAs ultrathin membrane, a tapered structure[44] is no longer required; such a structure requires a complicated fabrication process such as crystal regrowth to suppress the optical reflection at the edge of the Si hybrid waveguide[45,46]. As a result, the waveguide phototransistor shown in Fig. 1a can be realized with a simple fabrication process (see Methods and Supplementary Section I). Figure 1b shows a plan-view microscopy image of the fabricated photoFET. A 30-nm-thick p-type InGaAs layer lattice-matched to InP was bonded onto the Si waveguide with 10-nm-thick $Al_2O_3$ layer. Since the $Al_2O_3$ layer works as a gate dielectric, the electron inversion layer is formed when a positive gate voltage is applied through the Si waveguide, making the transistor turn on. Since the Schottky contact is formed between the p-type InGaAs and most metals owing to the Fermi level pinning[47], the metal source and drain were formed by simply depositing Ni/Au electrodes on the InGaAs layer. The channel length $L$ of the transistor, defined as the gap between the source and drain metals, was 2 μm. The length of the InGaAs absorber, corresponding to the width of the phototransistor $W$, was designed to be 30 μm. The Ni/Au contact to the Si waveguide back gate was formed on the heavily doped Si slab (see Supplementary Section I). The Si waveguide was designed to obtain a single-mode operation at a 1.3 μm wavelength. A light signal at a 1.3 μm wavelength was injected into the fundamental transverse-electric (TE) mode of the Si waveguide through a grating coupler and absorbed at the 30-nm-thick InGaAs layer.

First, the characteristics as a transistor without light injection were investigated (see Method). Figure 2a shows the drain current ($I_d$) – gate voltage ($V_g$) characteristics when the drain voltages ($V_d$) were 0.05 V and 0.5 V. By applying a gate voltage to the Si waveguide back gate, we obtained a drain current on/off ratio of approximately $10^4$, suggesting effective gate control for the transistor channel. Figure 2b shows the $I_d$–$V_d$ characteristics when $V_g$ was swept from -1 V to 1 V. The saturation of $I_d$ with respect to $V_d$ was observed, indicating a well-behaved transistor operation. We also prepared the InGaAs transistor for mobility measurement (see Supplementary Section II). The field-effect mobility was estimated to be 608 cm$^2$/V·s from the $I_d$–$V_g$ characteristics. The high electron mobility contributes to the high responsivity and the reasonable response time of the proposed InGaAs photoFET as discussed later.

Next, the characteristics of the photoFET were evaluated with light injection (See Method). Figure 3a shows the $I_d$–$V_d$ characteristics at $V_g$ of 1 V when the continuous-wave (CW) light signal at a 1305 nm wavelength was injected at various input powers. Note that the input power is defined as the intensity coupled to the photoFET by excluding the insertion loss of the grating coupler and the Si waveguide (see Supplementary Section III). Because of the photocurrent amplification by the transistor operation, a photocurrent significantly greater than the dark current was observed even for a very weak light of 631 fW. Figure 3b shows the $I_d$–$V_g$ characteristics in the linear scale at various input powers when $V_d$ was 0.5 V. It was clearly observed that the threshold voltage of the transistor was shifted by the photogating effect through hole accumulation in the InGaAs channel[37,39,48]. The photocurrent calculated by subtracting the dark current from the drain current with light irradiation is shown in Fig. 3c. A large photocurrent was obtained through the photogating effect when the transistor was in the on state. When the transistor was in the off state, the photoconductive effect might contribute to the amplification of the photocurrent as well as the photogating effect. Figure 3d shows the responsivity $R$ as a function of the input optical power $P$ at various $V_g$, where $R$ is proportional to $P^{\alpha-1}$. When $\alpha = 1$, the photoconductive effect is dominant, whereas when $0 < \alpha < 1$, the photogating effect is dominant[37,48–50]. It is found that $\alpha$ decreases as $V_g$ increases, suggesting that the photogating effect is more dominant in the on state. The photocurrent component attributable to the photogating effect can be expressed by the product of the transconductance and the threshold voltage shift. Since the transconductance increases as $V_g$ increases, the photocurrent amplified by the photogating effect becomes very large when the phototransistor is in the on state. As a result, when $V_g$ was 1 V, an extremely large responsivity of $2.1 \times 10^6$ A/W was obtained at an incident power of 631 fW.

The time response of the photoFET was also evaluated by injecting a modulated optical signal (See Method). The rise time and fall time of the optical signal generated by direct modulation of a tunable laser were approximately 400 ns and 100 ns, respectively (see Supplementary Section IV). Figure 4a shows the time response of $I_d$ when $V_d = 0.5$ V and $V_g = 1$ V. The rise time of $I_d$ ($\tau_R$) was approximately 1 μs, whereas the fall time ($\tau_F$) was approximately 100 μs, much longer than $\tau_R$. In contrast, when $V_g$ was -1 V, $\tau_R$ did not change significantly, whereas $\tau_F$ became much shorter (Fig. 4b). The gate voltage dependence of $\tau_R$ and $\tau_F$ is shown in Fig. 4c. Although $\tau_R$ was almost independent of $V_g$, $\tau_F$ increased exponentially with $V_g$. When the input light was turned off, the photocurrent induced by the photogating effect decreased owing to the reduction in the number of accumulated holes in the InGaAs channel through carrier recombination. As discussed for Fig. 3d, the photogating effect becomes dominant as $V_g$ increases. As a result, the hole lifetime makes $\tau_F$ long. In contrast, when the photoconductive effect is dominant with a negative $V_g$, the accumulated holes are more easily swept out from the channel through quantum tunneling, resulting in a short $\tau_F$.

Figure 5 is the benchmark for phototransistors that shows the relationship between responsivity and response time. Owing to the large optical absorption in the waveguide structure, the proposed waveguide-coupled photoFET exhibits more than three orders of magnitude greater responsivity than the surface-illuminated InGaAs photoFET[32]. PhotoFETs based on 2D materials[35,37,38,42] show a high responsivity of $10^3 - 10^7$ A/W. However, their response time is on the order of one second because of the long carrier lifetime of trap states, which is more suitable for image and gas sensor applications. Phototransistors based on BJTs or JFETs exhibit a response time of less than 1 ns, while their responsivity is less than $10^2$ A/W, which is suitable for high-speed optical communication applications. In contrast, using the InGaAs ultrathin channel and the Si waveguide gate, our device exhibits the highest responsivity of $10^5 - 10^6$ A/W among the waveguide-coupled phototransistors with a sub-millisecond response time, which makes it very suitable for an optical power monitor in Si photonic circuits. The photocurrent induced by the photogating effect ($I_{PG}$) can be expressed as

$$I_{PG} = g_m \Delta V_{th} = \frac{\Delta Q}{\tau_{tr}} = \frac{g^* \tau_p}{\tau_{tr}}, \qquad (1)$$

where $g_m$ is the transconductance, $\Delta V_{th}$ is the threshold voltage shift by the photogating effect, $\Delta Q$ is the total charge of the photogenerated holes, $\tau_{tr}$ is the carrier transit time of an electron[48], $g^*$ is the generation rate of photoexcitation, and $\tau_p$ is the carrier lifetime of holes. The proposed device has a larger $\Delta Q$ than the surface-illuminated InGaAs photoFETs because of the large optical absorption in the waveguide configuration. Since the carrier transit time $\tau_{tr}$ can be small owing to the high electron mobility in the InGaAs channel, we can increase $I_{PG}$ without increasing $\tau_p$. This is the reason why the proposed InGaAs photoFET can achieve a shorter response time than the phototransistors based on 2D materials with comparable responsivity. Since the electron mobility can be improved by process optimization due to the high electron mobility of InGaAs[47], there is room for improvement in the responsivity without the expense of the response time.

In conclusion, we successfully demonstrated the waveguide-coupled photoFET by bonding the InGaAs ultrathin membrane onto the Si waveguide. The InGaAs ultrathin channel enhanced the photogating effect through the Si waveguide gating, resulting in the highest responsivity among the waveguide-coupled phototransistors. The high responsivity and the response time of our phototransistor are sufficient for optical power monitoring and are close to the performance of a commercially available optical power monitor instrument. Owing to the high responsivity, the proposed phototransistor can be a transparent in-line optical power monitor for a Si waveguide realized by reducing the length of the InGaAs absorber to less than 1 μm (see Supplementary Section V). Hence, the waveguide-coupled InGaAs photoFET with a Si waveguide gate can be used as an effective optical power monitor in Si programmable photonic circuits for communication, computing, and sensing applications.

**Methods**

The InGaAs photoFET was fabricated as follows (see Supplementary Section I for the detailed fabrication procedure). A 30-nm-thick p-type $In_{0.53}Ga_{0.47}As$ membrane was bonded onto the Si waveguide with 10-nm-thick $Al_2O_3$ bonding interface. After the patterning of the InGaAs membrane, Ni/Au contact pads were formed as the source, drain, and gate by lift-off. The electrical characteristics of the photoFET were measured using a semiconductor parameter analyzer (Agilent Technologies, 4156C). The photoresponse of the photoFET was measured at a wavelength of 1305 nm. A tunable laser (Santec, TSL-

510) was used as a fiber-coupled light source. The input power was tuned using a variable optical attenuator (Anritsu, MN9605C). The polarization of the input light was adjusted to the TE mode of the Si waveguide by an in-line polarization controller. The input light was coupled from a single-mode fiber to the Si waveguide through the grating coupler. For the time response measurement, the tunable laser was modulated using an electrical waveform generator (Agilent, 33522B). The electrical waveform was recorded using a waveform generator/fast measurement unit (Agilent, B1530A) of a semiconductor device analyzer (Agilent Technologies, B1500A).

**Acknowledgements**

This work was partly based on results obtained from projects (JPNP14004, JPNP16007) commissioned by the New Energy and Industrial Technology Development Organization (NEDO) and partly supported by JST, CREST Grant Number JPMJCR2004. Part of this work was conducted at Takeda Sentanchi super cleanroom, The University of Tokyo, supported by "Nanotechnology Platform Program" of the Ministry of Education, Culture, Sports, Science and Technology (MEXT), Japan, Grant Number JPMXP09F20UT0021. The authors thank Dr. H. Yagi, Dr. Y. Itoh, and Dr. H. Mori of Sumitomo Electric for providing InP epitaxial wafers.

**Author contributions**

T.O. contributed to fabrication, measurement, and manuscript preparation. K.S. and S.O. contributed to the fabrication and analysis. S.M. and F.B. contributed to design and fabrication of the Si waveguide. K.T. and S.T. contributed to the overall discussion. M.T. contributed to the idea, discussion, and manuscript revision and also provided high-level project supervision.

**Competing financial interests**

The authors declare no competing financial interests.

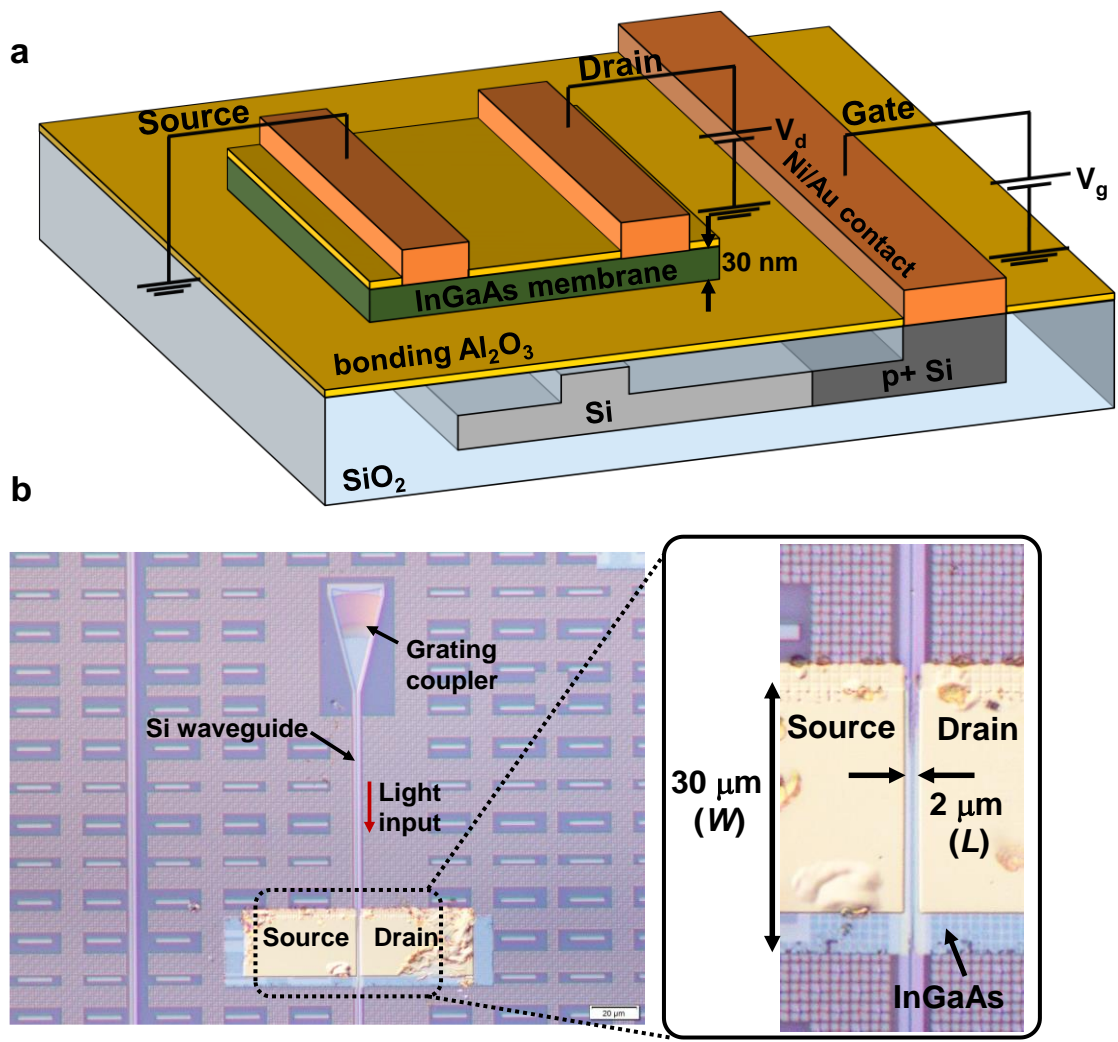

Fig. 1. **Schematic and images of a fabricated waveguide-coupled phototransistor. a,** 3D view of the phototransistor consisting of a 30-nm-thick InGaAs membrane on the Si waveguide back gate with an Al$_2$O$_3$ gate dielectric. The Ni/Au metal source and drain are formed on the InGaAs channel. **b,** Plan-view microscopy images of the fabricated device. An optical signal at a 1.3 μm wavelength is coupled to the Si waveguide through the grating coupler. The gate length and width are 2 μm and 30 μm, respectively.

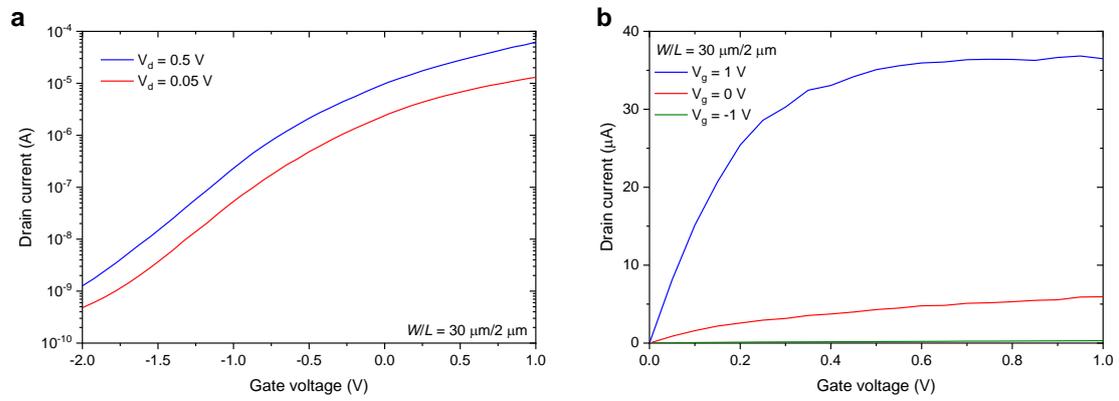

Fig. 2. **Electrical characteristics of a waveguide-coupled InGaAs phototransistor with no light injection. a,** $I_d$–$V_g$ characteristics with $V_d$ of 0.05 V and 0.5 V. **b,** $I_d$–$V_d$ characteristics with $V_g$ of -1 V, 0 V, and 1 V.

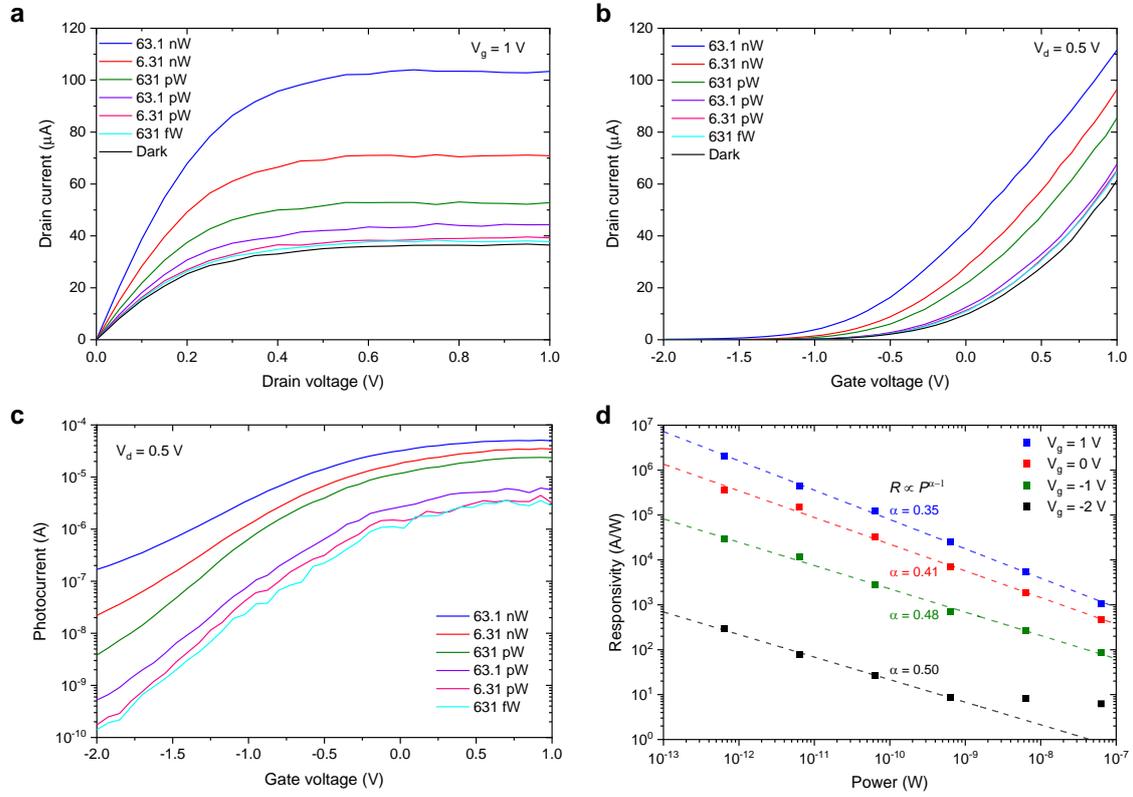

Fig. 3. **Photoresponses of a waveguide-coupled InGaAs phototransistor with light injection of a 1.3 μm wavelength. a,** $I_d$–$V_d$ characteristics with $V_g$ of 1 V. **b,** $I_d$–$V_g$ characteristics in the linear scale with $V_d$ of 0.5 V. **c,** Photocurrent as a function of $V_g$ with $V_d$ of 0.5 V. **d,** Relationship between responsivity and input power at various gate voltages.

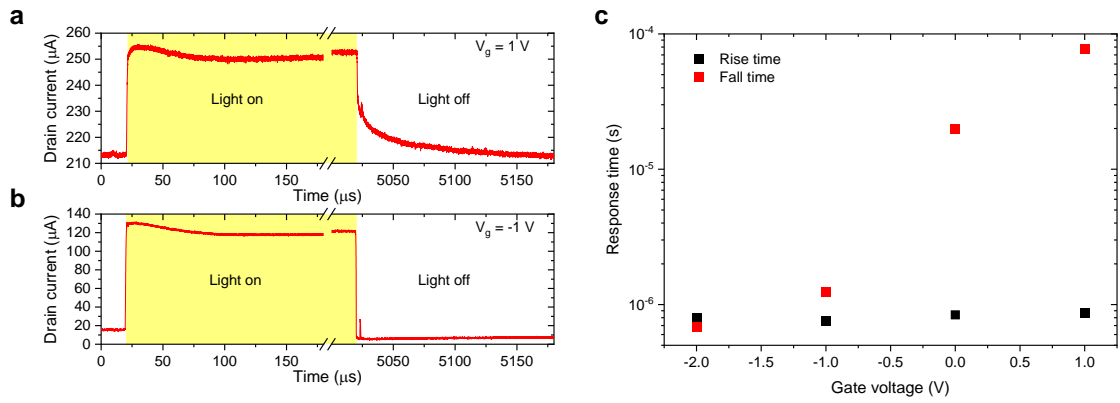

Fig. 4. **Time response of a waveguide-coupled InGaAs phototransistor. a,** Time response of $I_d$ with $V_g$ of 1 V. **b,** Time response of $I_d$ with $V_g$ of -1 V. **c,** Rise time and fall time as a function of $V_g$.

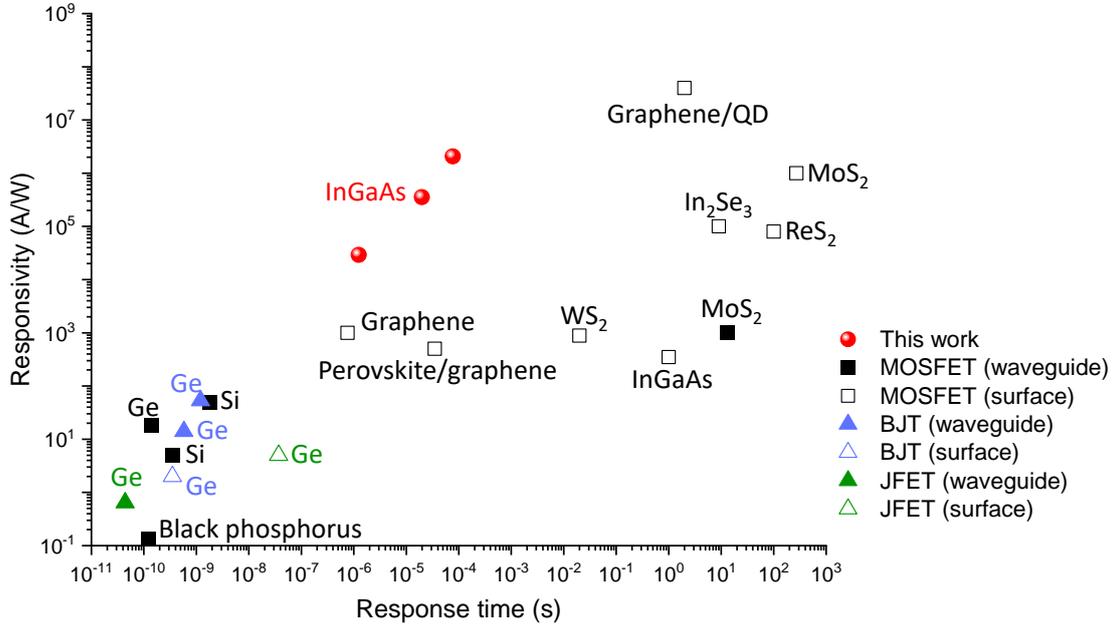

Fig. 5. **Benchmark of responsivity and response time for various phototransistors based on metal-oxide-semiconductor field-effect transistors (MOSFETs), bipolar junction transistors (BJTs), and junction field-effect transistors (JFETs).** Comparison of responsivity and response time demonstrated for MOSFET-based phototransistors (waveguide-coupled: Ge[31], Si[43], black phosphorus[29] and MoS$_2$[42], surface-illuminated: InGaAs[32], graphene/QD[34], perovskite/graphene[40], graphene[39], MoS$_2$[35], WS$_2$[41], In$_2$Se$_3$[37], and ReS$_2$[38]), BJT-based phototransistors (waveguide-coupled: Ge[22,24], surface-illuminated: Ge[25]), and JFET-based phototransistors (waveguide-coupled: Ge[27], surface-illuminated: Ge[28]).

# Supplementary Information

## Ultrahigh-sensitivity optical power monitor for Si photonic circuits

Takaya Ochiai[1], Kei Sumita[1], Shuhei Ohno[1], Stéphane Monfray[2], Frederic Boeuf[2], Kasidit Toprasertpong[1], Shinichi Takagi[1], Mitsuru Takenaka[1*]

[1] Department of Electrical Engineering and Information Systems, The University of Tokyo, 7-3-1 Hongo, Bunkyo-ku, Tokyo 113-8656, Japan,

Phone: +81-3-5841-6733, Fax: +81-3-5841-8564,

[2] STMicroelectronics, 850 Rue Jean Monnet 38920 Crolles, France

*E-mail: takenaka@mosfet.t.u-tokyo.ac.jp

## Contents

## I.   Fabrication procedure for waveguide-coupled InGaAs phototransistor

Figure S1 shows the process flow of the waveguide-coupled InGaAs phototransistor. Si rib waveguides were fabricated on a Si-on-insulator (SOI) wafer with a 300-nm-thick Si layer. The width and the rib height were 400 nm and 150 nm for a single-mode operation at a 1.3 μm wavelength. Grating couplers for the transverse electric (TE) mode were integrated with the Si waveguide for fiber coupling. After forming $SiO_2$ cladding by chemical vapor deposition (CVD) on the Si waveguide, chemical mechanical polishing (CMP) was carried out for surface planarization. To form the p-Si (acceptor concentration $N_A = 5 \times 10^{17}$ cm$^{-3}$) and p$^+$-Si ($N_A = 1 \times 10^{20}$ cm$^{-3}$) regions, boron ions were implanted. Then, an InP epitaxial wafer consisting of a p-type 30-nm-thick In$_{0.53}$Ga$_{0.47}$As layer ($N_A = 5 \times 10^{16}$ cm$^{-3}$) and etch-stop layers was bonded onto the Si waveguide with an Al$_2$O$_3$ bonding layer formed by atomic layer deposition (ALD)[1]. After bonding, the InP wafer and the etch-stop layers were selectively removed by wet etching. The 40-nm-thick SiO$_2$ hard mask formed by CVD was patterned by electron-beam (EB) lithography and inductively coupled plasma (ICP) etching. Then, the InGaAs layer was patterned by wet etching. After removing the SiO$_2$ hard mask, a 10-nm-thick Al$_2$O$_3$ passivation layer was formed by ALD. Finally, the contact windows for source and drain (S/D) were opened, and Ni/Au contact pads as metal S/D[2,3] were formed by EB evaporation and lift-off. The Ni/Au contact pad for the gate was also formed simultaneously. Figure S2 shows a plan-view microscopy image of the fabricated device. The contact pad for the Si waveguide gate was formed away from the S/D region owing to the limitation of the original Si photonic circuit design.

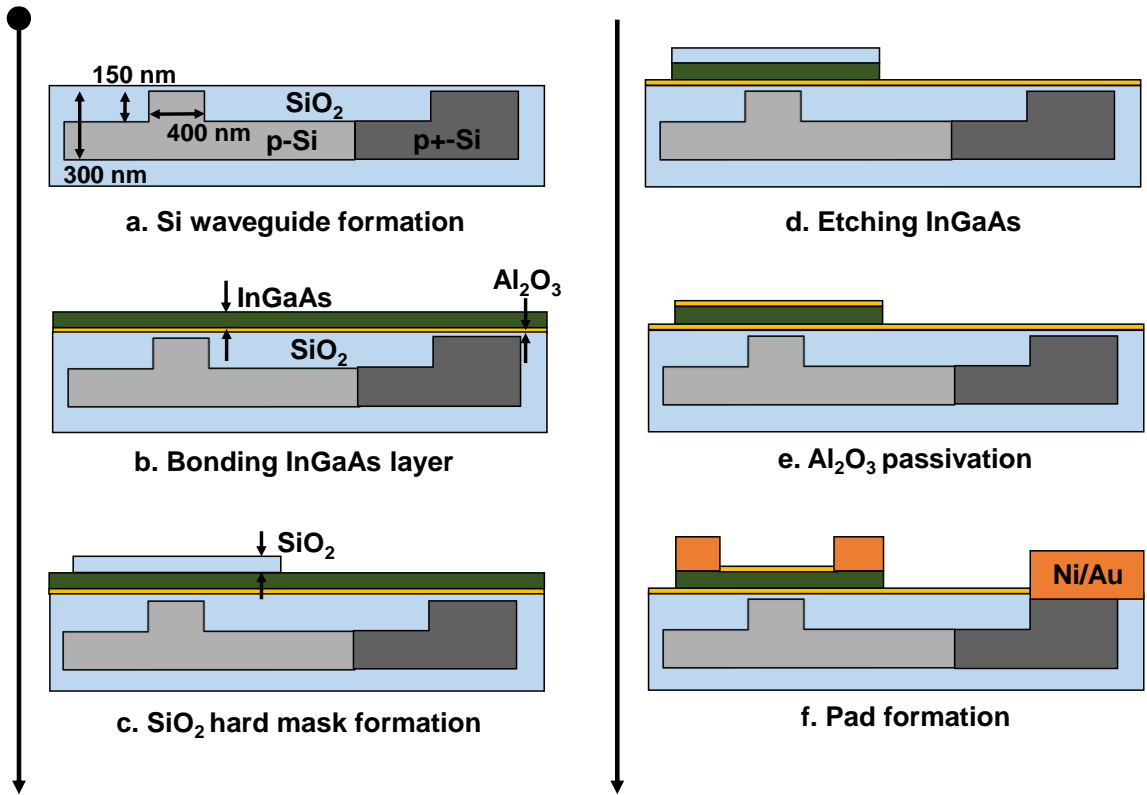

Fig. S1. **Process flow of waveguide-coupled InGaAs phototransistor.** An InGaAs ultrathin membrane is bonded onto a p-type Si waveguide with $Al_2O_3$ gate dielectric. The Si waveguide acts as a gate electrode, and the Ni/Au metal source and drain are deposited on the InGaAs channel.

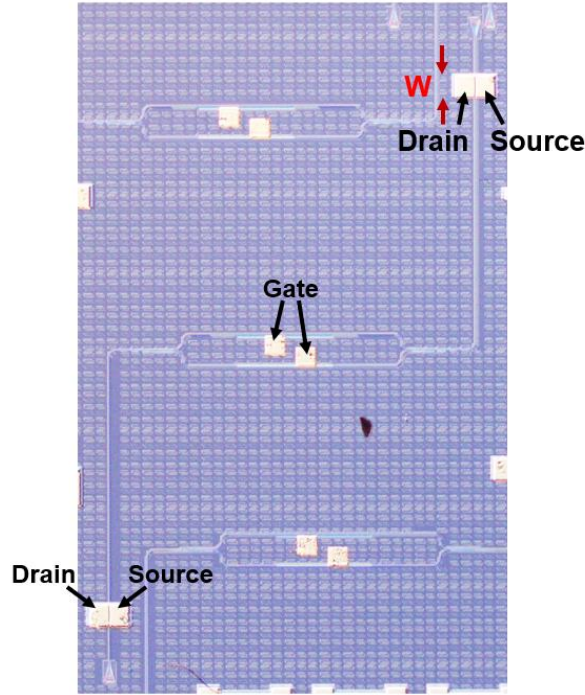

Fig. S2. **Plan-view microscopy image of waveguide-coupled InGaAs phototransistors.**

## II. Measurement of electron mobility of InGaAs transistor

We prepared the InGaAs transistor shown in Fig. S3 for mobility measurement. From the drain current ($I_d$) – gate voltage ($V_g$) characteristics in Fig. S4a, we extracted the field-effect electron mobility $\mu_{FE}$ as

$$\mu_{FE} = \frac{dI_d}{dV_g}\frac{L}{W}\frac{1}{C_{ox}V_d},$$

where $L$ is the channel length, $W$ is the channel width, and $C_{ox}$ is the gate capacitance. Here, $L$ was 20 μm, and $W$ was 30 μm. $C_{ox}$ was calculated to be 0.298 μF/cm$^2$ by taking into account the 6-nm-thick SiO$_2$ and 10-nm-thick Al$_2$O$_3$ layers. Figure S4b shows the extracted $\mu_{FE}$ as a function of $V_g$. The peak value of the electron mobility was approximately 608 cm$^2$/Vs. Note that there is room for improvement in electron mobility through the process optimization due to the high electron mobility of InGaAs[3].

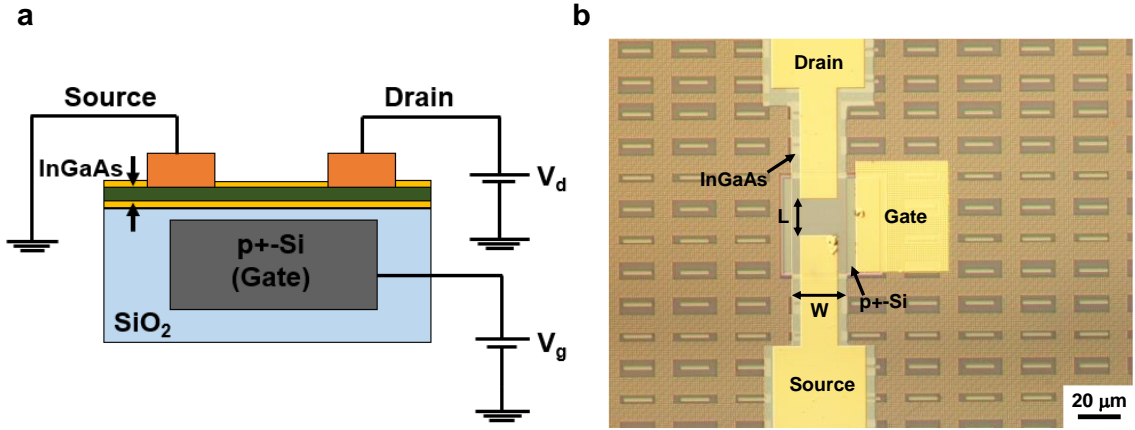

Fig. S3. **InGaAs transistor for mobility measurement. a,** Cross-sectional schematic of the InGaAs transistor. **b,** Plan-view microscopy image of the InGaAs transistor.

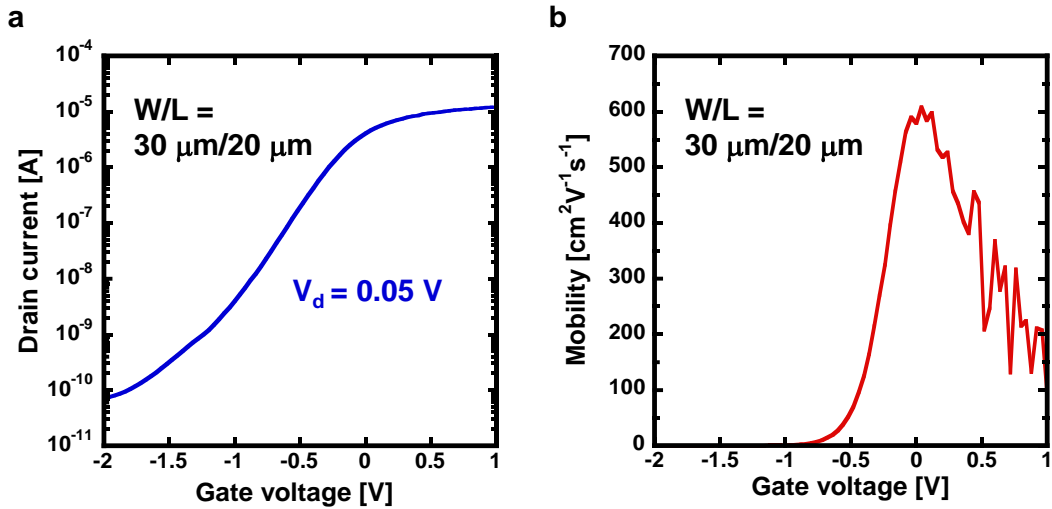

Fig. S4. **Electrical characteristics of the InGaAs transistor. a,** Drain current ($I_d$) – gate voltage ($V_g$) characteristics. **b,** Field-effect electron mobility as a function of gate voltage.

## III. Propagation loss of the Si waveguide

We evaluated the propagation loss of the Si waveguide and the coupling loss of the grating coupler at a wavelength of 1305 nm to obtain the optical power injected into the phototransistor. We prepared the Si waveguide of various waveguide lengths from 1296 μm to 12251 μm, as shown in Fig. S5a. A continuous wave (CW) light from a tunable laser was coupled into the Si waveguide through the grating coupler, and the output power from the Si waveguide was coupled again to a single-mode fiber through another grating coupler. The output power was measured using an InGaAs optical power monitor. The polarization of the input light was tuned to the transverse-electric mode using a polarizer. A variable attenuator was also inserted between the tunable laser and the polarizer to adjust the input power. The insertion loss from the tunable laser to the power meter without the device under test was measured to be 5.2 dB. Figure S5b shows the transmission spectra of the Si waveguide of various waveguide lengths. From the results in Fig. S5b, the propagation loss of the Si waveguide and the coupling loss of the grating coupler were extracted to be 2.14 dB/cm and 7.5 dB, respectively, as shown in Fig. S5c.

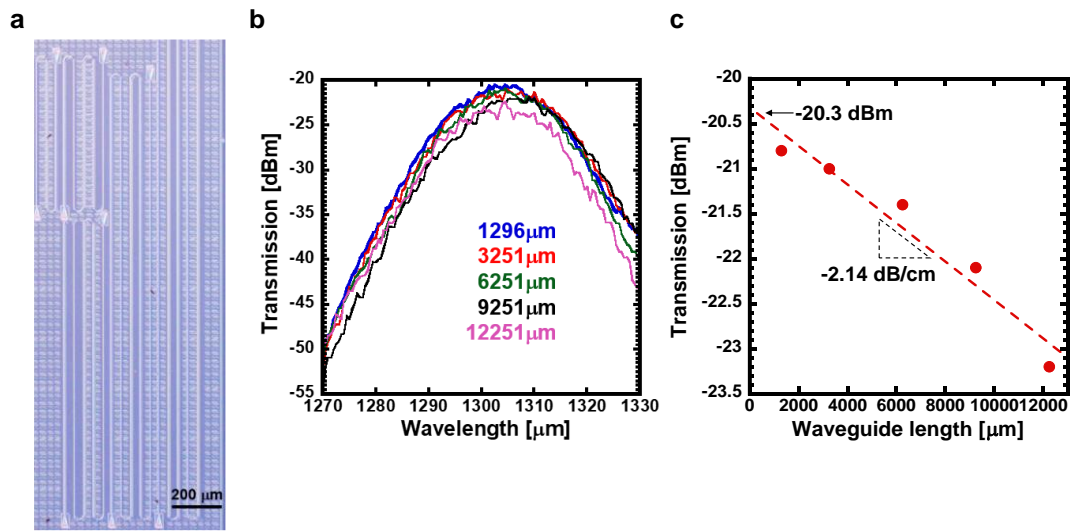

Fig. S5. **Measurement of the propagation loss and coupling loss. a,** Plan-view microscopy image of the Si waveguides of various waveguide lengths for loss measurement. **b,** Transmission spectra of the Si waveguide of various waveguide lengths. **c,** Transmission as a function of waveguide length.

**IV. Waveform of the modulated light**

The tunable laser was modulated with a 100-Hz electrical square signal from a function generator to generate modulated light for time response measurement. To confirm the rise time and fall time of light modulation, the waveform of the modulated light was measured using a photodetector with the response time of approximately 20 ns, as shown in Fig. S6. The rise time and fall time were 400 ns and 100 ns, respectively.

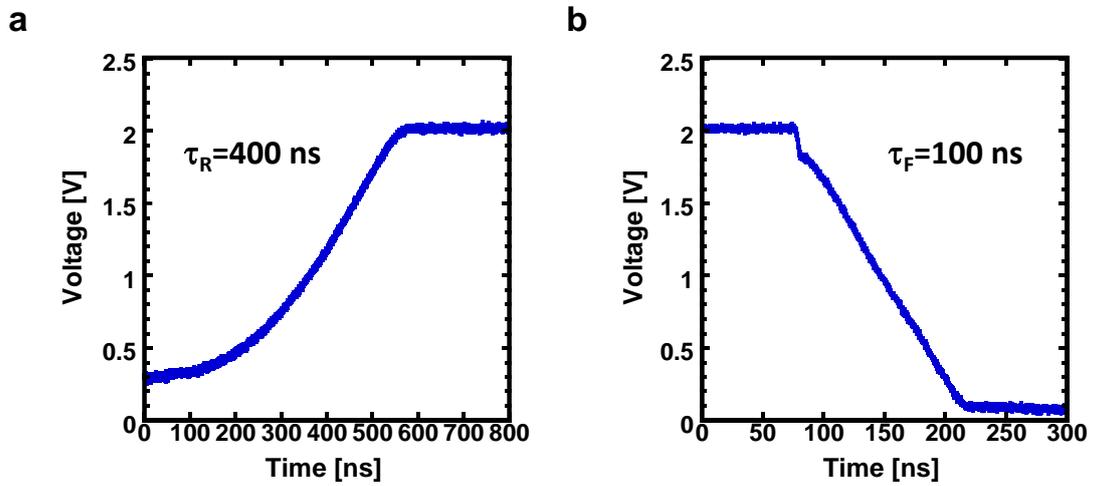

Fig. S6. **Waveform of the modulated light. a,** Rising edge. **b,** Falling edge.

**V. Insertion loss of the waveguide-coupled phototransistor**

The insertion loss of the 30-µm-long InGaAs phototransistor was 7 dB, taking into account the propagation loss of the Si waveguide and the coupling loss of the grating coupler. Figure S7 shows the expected insertion loss of the InGaAs phototransistor as a function of device length. Since the insertion loss per unit length is 0.23 dB/µm, we expect that the insertion loss can be less than 0.1 dB with the 435-nm-long phototransistor. As the device length decreases, the dark current also decreases. In addition, the total number of holes for achieving the same threshold voltage shift also decreases since the volume of the channel decreases with the device length. Thus, the sensitivity of the scaled phototransistor is expected not to degrade markedly. Because of the high responsivity of the phototransistor, we can use it as an optical power monitor even with the device length of much smaller than 435 nm. Therefore, the proposed waveguide-coupled InGaAs phototransistor can potentially be used as a transparent optical power monitor for a Si waveguide.

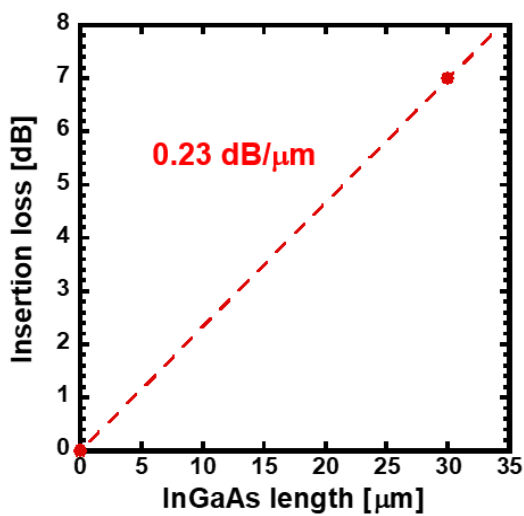

Fig. S7. **Insertion loss of the phototransistor as a function of InGaAs length.**

source/drain. *Appl. Phys. Express* **4**, 114201 (2011).